\documentclass[a4paper]{jpconf}
\usepackage{graphicx}
\begin{document}
\title{Dynamical bottomonium-suppression in a realistic AA background}

\author{P-B Gossiaux and R Katz}

\address{Subatech, 4 rue Alfred Kastler, 44307 Nantes, France}

\ead{gossiaux@subatech.in2p3.fr, katz@subatech.in2p3.fr}

\begin{abstract}
We refine our dynamical Schroedinger-Langevin scheme designed for dealing with the suppression of quarkonium states in URHIC in order to allow for feed-downs from higher states and present updated results for the $R_{AA}$ as a function of the centrality.
\end{abstract}

\section{Introduction}
The suppression of upsilon $\Upsilon(1S)$ states in AA collisions, observed by the STAR collaboration at RHIC~\cite{STAR:2014} and by the CMS~\cite{CMS:2015} and ALICE~\cite{ALICE:2014} collaborations at LHC, is one of the most convincing evidence for the creation of the quark gluon plasma. The precise survival of excited $\Upsilon(2S)$ and $\Upsilon(3S)$ states vs ground state could even allow to measure the highest temperature reached in those collisions, according for instance to the sequential suppression scenario which is substantiated by calculations of the dissociation temperature based on lattice potentials evaluated at finite temperature. In a previous work~\cite{Gossiaux_Katz:2016}, we have addressed the question of upsilon dissociation resorting to a dynamical approach, i.e. the non-linear Schroedinger-Langevin equation (SLE). In this scheme, the time-dependent real potential implements the Debye-screening while the stochastic forces express the (hard) interactions between the QGP and the $b\bar b$ state. The SLE allows to treat transitions between bound and open quantum states (the "suppression"), but also transitions between bound states, a feature usually lacking in other treatments. In a stationary QGP, SLE naturally leads to asymptotic distributions of {Y(1S), Y(2S),...} following correct statistical weights~\cite{Katz:2015qja}, which allows to make the link with models based on the hypothesis of statistical recombination. This sanity check is another appealing feature of our approach.

In this work, we iterate on our previous predictions \cite{Gossiaux_Katz:2016} for upsilon suppression in URHIC at LHC energy. In order to achieve 
more reliable comparisons between our model and experimental measurements, two improvements were implemented, as described in sections \ref{section_model} and \ref{section_productionpp}.

\section{The model}
\label{section_model}

As detailed in \cite{Gossiaux_Katz:2016}, we resort to the so-called Schr\"odinger-Langevin equation (SLE) to deal with the dynamics of the $b-\bar{b}$ internal
state:
\begin{equation}
i\hbar\frac{\partial \psi}{\partial t}= \Bigg[
 \hat{H}_0 + \hbar A\Big(S({\bf x},t)-\int\psi^*S({\bf x},t)\,\psi \,\,d{\bf x }\Big) -
{\bf x}\cdot{\bf F}_R(t)\Bigg]\,\psi\,
\;\;{\rm with}\;\;
\hat{H}_0 =-\frac{\hbar^2 \Delta}{m_b}+V_{\rm MF}({\bf x})\,.
\label{SLeq}
\end{equation}
In this equation, the non linear term $\propto \hbar A$ is a dissipative friction term based on the wave function phase $S$ while the term $-{\bf x}\cdot{\bf F}_R(t)$ represents stochastic dipolar forces mocking the scatterings on QGP constituents. The Hamiltonian $\hat{H}_0$ encodes the dynamics in the absence of such stochastic interactions with the medium, resorting to the mean field potential $V_{\rm MF}$. 

Solving (\ref{SLeq}) in 3+1D is still beyond our present implementation of the numerical scheme. Consequently, we adopt a 1+1D modeling of the $b\bar{b}$ pairs, with even (resp.~odd) 1D-states mocking S (resp.~P) 3D-states. In vacuum, the potential in $\hat{H}_0$ is taken as $V_{\rm MF}(T=0,x)=V_{\rm vac}(x)=K|x|$, truncated to $V_{\rm max}$. Such an approximation was already adopted in \cite{Gossiaux_Katz:2016}, with (set I) $m_b=4.575~{\rm GeV}$, $K = 1.375~{\rm GeV/fm}$ and 
$V_{\rm max}=1.2~{\rm GeV}$, following lQCD calculations~\cite{Mocsy:2008}. The masses of the various upsilon states were found in good agreement with their experimental values but the binding energies a bit too low
(in particular for $\Upsilon(3S)$, whose binding energy was just 20 MeV in our model). In this work, we use slightly different values of the parameters, i.e. (set II) $m_b=4.61~{\rm GeV}$, $K=1.246~{\rm GeV/fm}$ and $V_{\rm max}=1.338~{\rm GeV}$, which leads to very good values of the bottomonia masses ($E_{1S}=9.46~{\rm GeV}$, $E_{1P}=9.77~{\rm GeV}$, $E_{2S}=9.99~{\rm GeV}$, $E_{2P}=10.18~{\rm GeV}$, $E_{3S}=10.35~{\rm GeV}$) and even leads to a 3P state (not found for the set I), with  $E_{3P}=10.51~{\rm GeV}$. With set II, the binding energy of the 3S state is 120~{\rm MeV}, which is the correct value in the vacuum.

\section{Production in pp}
\label{section_productionpp}
\begin{figure}[h!]
\includegraphics[scale=0.48]{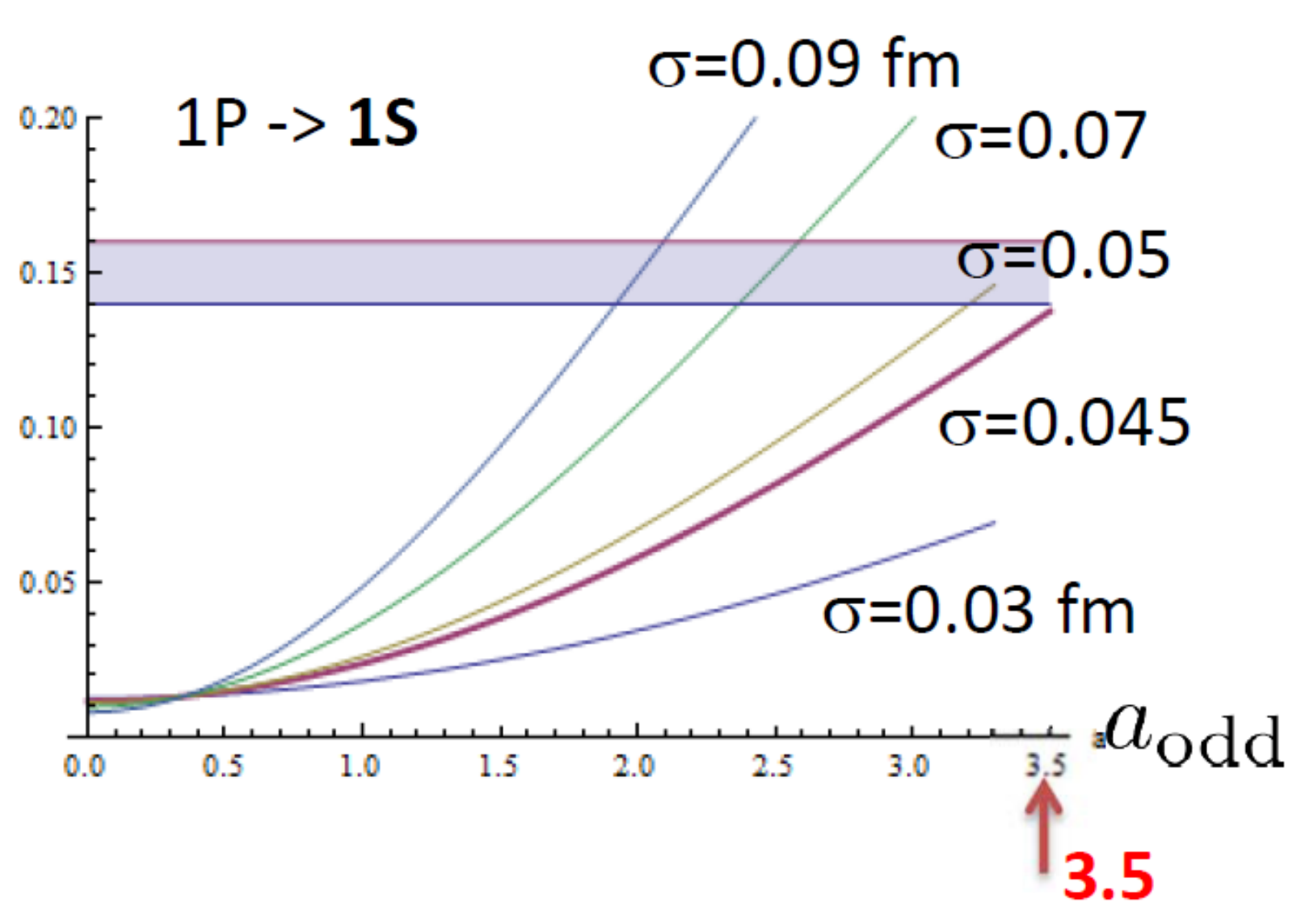}
\hspace{0.5cm}
\includegraphics[scale=0.65]{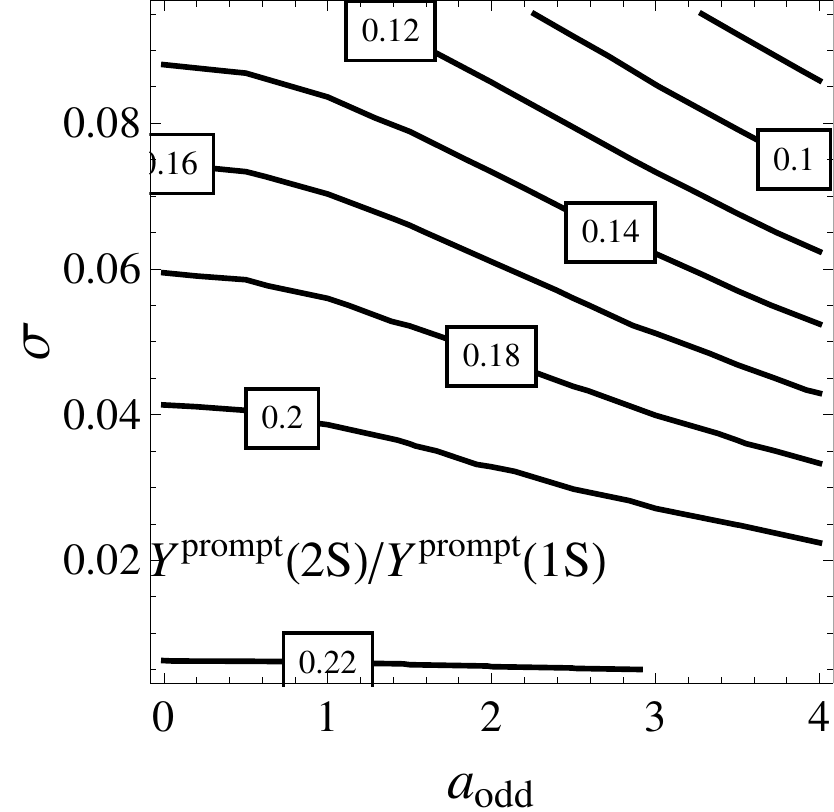}
\caption{Left panel: proportion from the prompt 1S-bottomonia production stemming from 1P decay as a function of $a_{\rm odd}$ for various $\sigma$ values. The band represent the 
experimental result at small $p_T$. Right panel: 
$\Upsilon(2S)/\Upsilon(1S)$ ratio as a function of $\sigma$ and $a_{\rm odd}$.}
\label{fig1}
\end{figure}

The main improvement of this contribution wrt our previous work consists in dealing with the feed-downs. For this purpose, we need a fair understanding of the direct production of the various bottomonium states within our scheme, starting from the pp case.
This is not a trivial issue, as the debate between various descriptions (Color Evaporation Model, Color Singlet Model, Color Octet Model) is still not settled~\cite{Andronic:2016}. Our {\it bona fide} prescription
is to consider some trial initial state $\psi_{b\bar{b}}(t=0)$ in the singlet representation containing even ($\equiv$ S wave) and
odd ($\equiv$ P wave) components:
\begin{equation}
 \psi_{b\bar{b}}(t=0,x) \propto 
 \left(1+ a_{\rm odd} \frac{x}{\sigma}\right)
 e^{-\frac{x^2}{2\sigma^2}}\,,
 \label{eq_init_state}
\end{equation}
where $\sigma$ and $a_{\rm odd}$ are free parameters. In pp, the direct production of $\Upsilon(nS)$ or of $\chi_b(nP)$ is then
modeled as $d\sigma^{\rm direct}_{\Upsilon(nS)/\chi_b(nP)}\propto 
|\langle nS/nP |  \psi_{b\bar{b}}(t=0)\rangle|^2$, where $|nS\rangle$ and $|nP\rangle$ are the eigenstates for $V_{\rm vac}$. The prompt production is then obtained from the branching ratios $\beta_{ij}$ through $d\sigma^{\rm prompt}_i=
d\sigma^{\rm direct}_i + \sum_{j>i} \beta_{ij} d\sigma^{\rm direct}_j$. This allows to deduce the various contributions from feed-downs and extract the optimal parameters $a_{\rm odd}$ and $\sigma$ by comparison with experimental 
results~\cite{Andronic:2016}. Let us insist that this procedure is pretty much specific to quantum treatments while semi-classical modeling like transport equations "just" rely on the probabilities
and can thus be initiated with the mere numbers extracted from measured $d\sigma^{\rm direct}$. The extraction of the optimal parameters was made considering the experimental values measured at small
$p_T$ which represents the bulk of the production. For the purpose of illustration we show in fig.~\ref{fig1} (left), the proportion of 1S prompt production stemming from 1P decay, as a function of both parameters. Rather unexpectedly, it appears to be 
possible to reproduce all  low-$p_T$  feed-downs ratios with the simple form (\ref{eq_init_state}) as initial state for $\sigma=0.045~{\rm fm}$ and $a_{\rm odd}=3.5$. With these parameters, the absolute (direct) weights in 1S, 1P, 2S, \ldots, 3P are 5.5\%, 3.6\%, 1.5\%, 3.3\%, 0.9\% and 2.5\% respectively.    It should be noted, however, that rather strong $p_T$ dependences of this "cocktail" has been reported in \cite{Andronic:2016}, f.i. an increase of the feed-down contribution from 1P$\rightarrow$ 1S (LHCb) as well as the increase of the (prompt) $\Upsilon(nS)/\Upsilon(1S)$ ratios (ATLAS, CMS), reaching $\sim 0.5$ for $p_T\sim 30~{\rm GeV}/c$ which, to our knowledge has not yet been explained on theoretical grounds.  

We have tried to accommodate these evolutions with the trial state
(\ref{eq_init_state}), varying the parameters $\sigma$ and $a_{\rm odd}$ with $p_T$ but we have identified that the (prompt) ratios $\Upsilon(2S/3S)/\Upsilon(1S)$ cannot exceed $\approx 0.22/0.16$
(as illustrated on fig. \ref{fig1} for the $\Upsilon(2S)/\Upsilon(1S)$ ratio) which are significantly smaller than the ratios found experimentally at large $p_T$. As a consequence, more sophisticated production mechanisms should be investigated and understood in this domain, for which our predictions are thus less robust.

\section{Production in PbPb and perspectives}
In PbPb collisions, the initial state (\ref{eq_init_state})
is evolved as in \cite{Gossiaux_Katz:2016}. The temperature and velocity
profile is modeled by EPOS2~\cite{EPOS:2010}, which provides good predictions in the soft sector, despite resorting to an ideal fluid dynamics evolution (this being somehow compensated by broader initial flux tubes).
In order to model the temperature dependence of the mean-field $V_{\rm MF}(T,x)$, we {\em bona fide} scale down $V_{\rm max}$ according to the asymptotic value of the
so-called "weak" potential $V_W$ introduced by Mocsy and Petreczky~\cite{Mocsy:2008} (our privileged choice in these proceedings), characterized by a binding strength intermediate between the free energy $F$ and the internal energy $U$~\cite{Kaczmarek:2005}. 
 
\begin{figure}[h!]
\hspace{1 cm}
\includegraphics[scale=0.25]{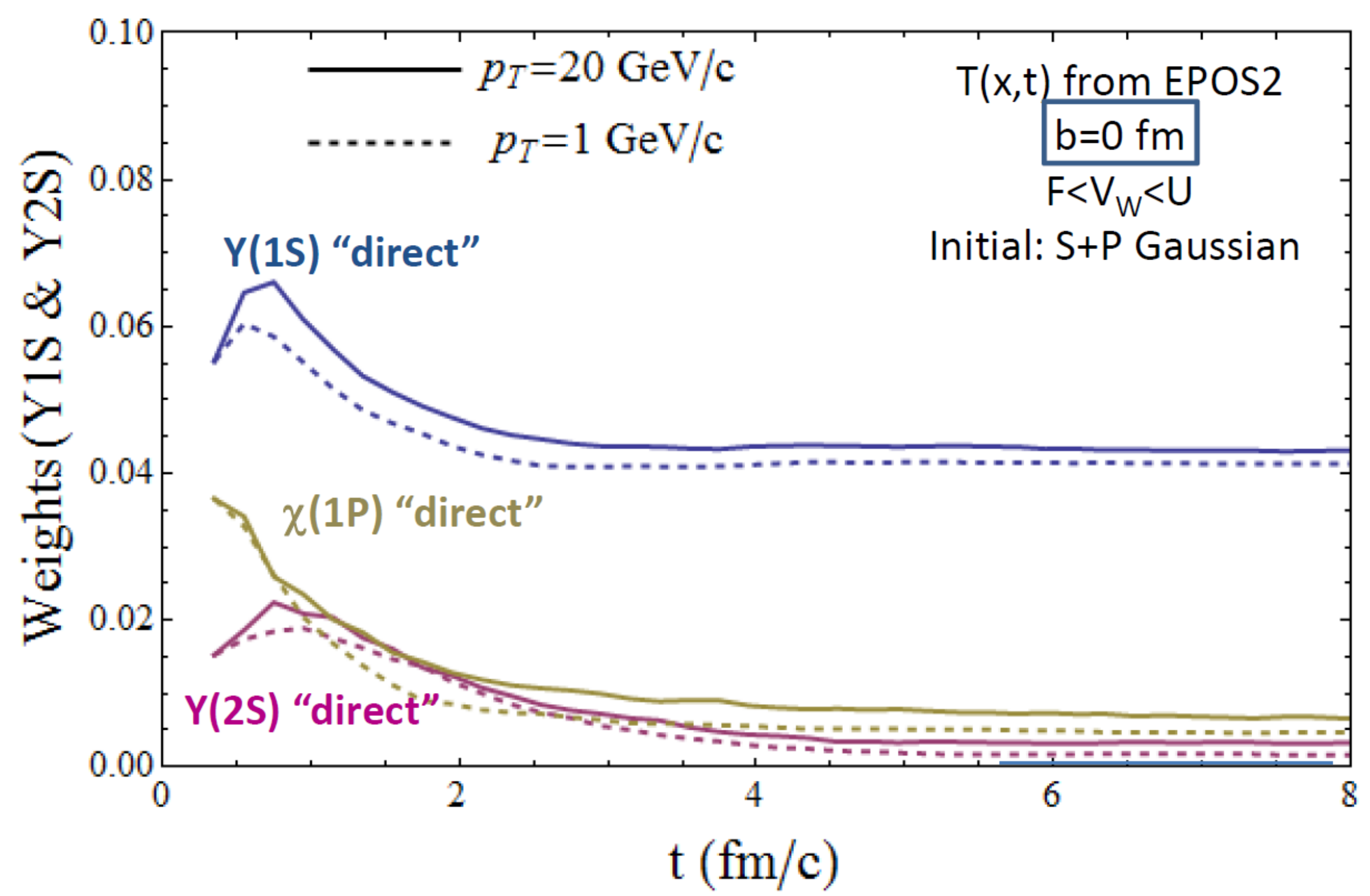}
\hspace{1 cm}
\includegraphics[scale=0.28]{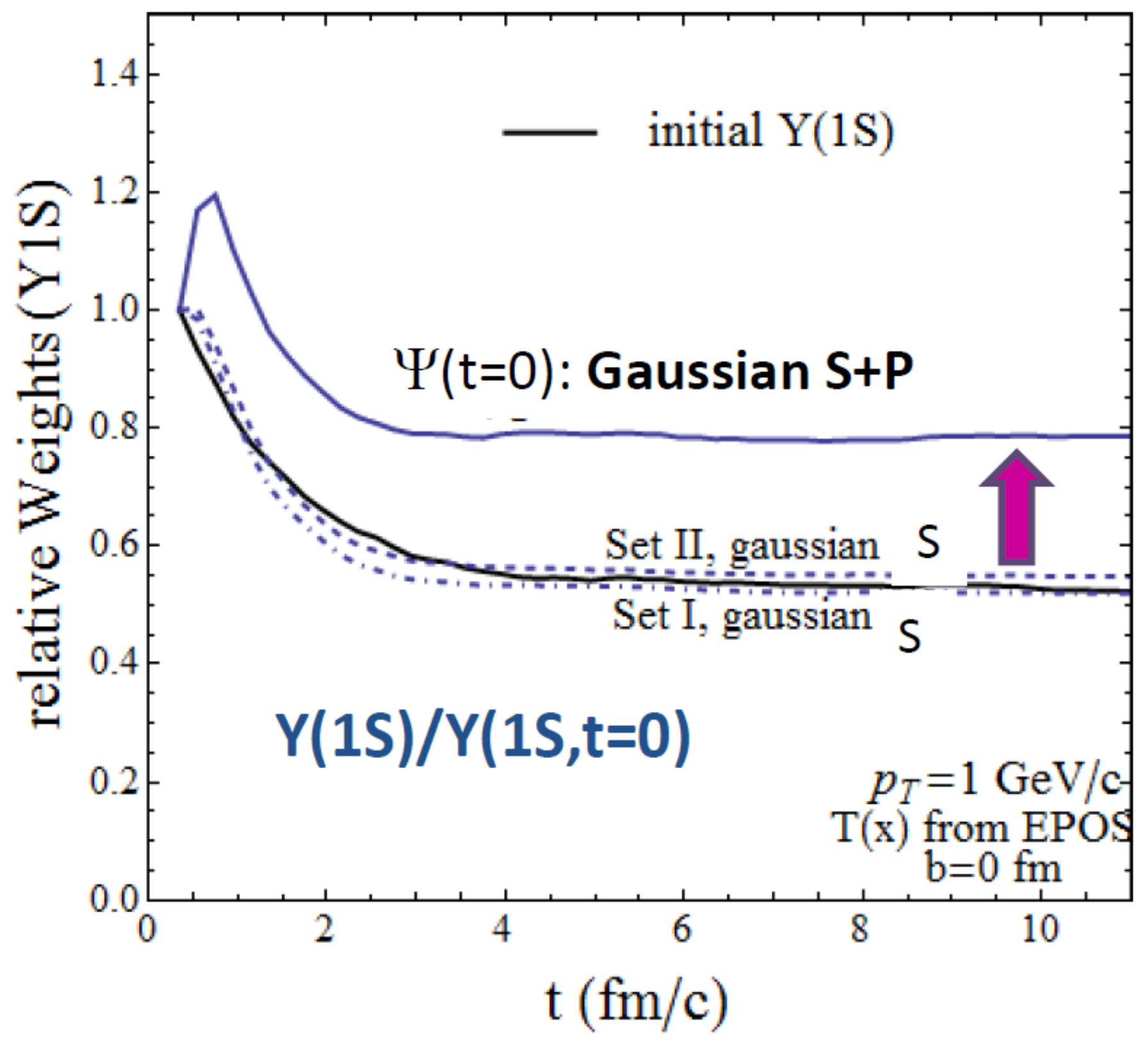}
\caption{Left panel: average evolution of the 1S, 1P and 2S components of the quantum state in the EPOS2 temperature profile for a $b\bar{b}$ state produced according to an S+P initial Gaussian state with $p_T=1~{\rm GeV/c}$ (dashed)
and $p_T=20~{\rm GeV/c}$ (solid). Right panel: "survival" of
the 1S state for various models.}
\label{fig2}
\end{figure}

In fig.~\ref{fig2}, we show the time evolution of the different components of the quantum state, after averaging on a) stochastic forces, b) production point and initial direction, and c) EPOS2 events, for a central PbPb collision at $\sqrt{s}=2.76~A\,{\rm GeV}$ and 2 choices of $p_T$. Quite generally, one observes a strong suppression of excited states, whose weights evolve on longer time scales. We also notice that the 
S-states benefit from some extra population from the P states during the first fm/c, due to the dipolar nature of the stochastic forces, a feature lacking in \cite{Katz:2015qja} where an initial S-state corresponding to $a_{\rm odd}=0$ was chosen. The $p_T$ dependence is found to be mild. In the right panel of  fig.~\ref{fig2}, we show the "survival" of $\Upsilon(1S)$ -- i.e. the instantaneous 1S-component of the state relative to the initial one -- for various models. We see that this survival is not much affected by the new set of parameters, neither by the specific wave packet in the initial S wave (Gaussian of 1S-eigenstate), while starting from mixed S+P wave packet leads to a larger survival (of the order of 80\% instead of 60\%), mostly due to the initial P$\rightarrow$ S transitions.   

\begin{figure}[h!]
\hspace{1 cm}
\includegraphics[scale=0.3]{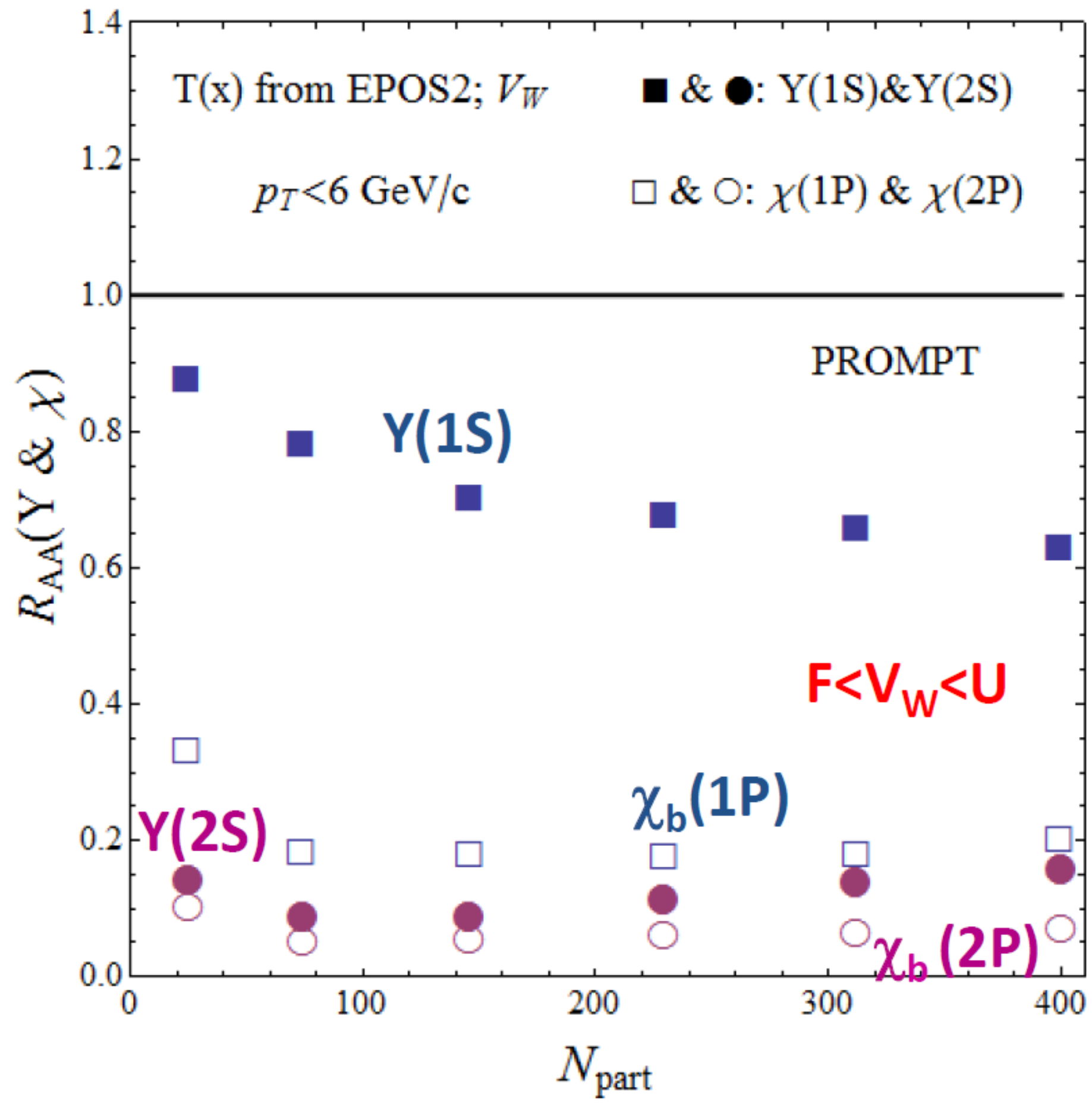}
\hspace{1 cm}
\includegraphics[scale=0.3]{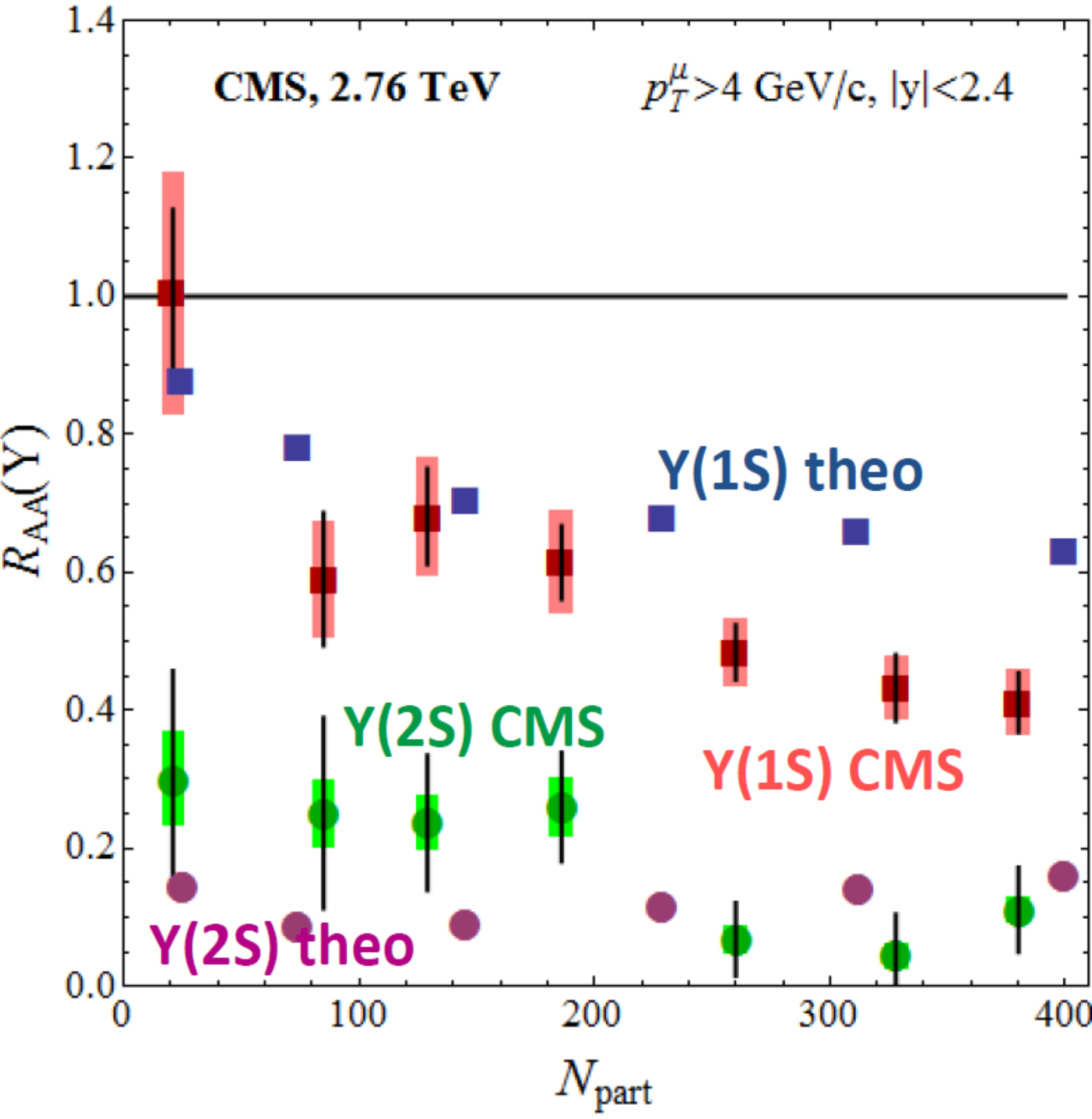}
\caption{Left panel: $R_{PbPb}$ of various bottomonium states as a function of the number of participants in the collision. Right panel: same for the upsilon states, compared with CMS data~\cite{CMS:2015}.}
\label{fig3}
\end{figure}
In fig.~\ref{fig3}, we show the nuclear modification factor for the prompt production of all bottomonium-like states in our model, applying the suppression of the direct states before feed-downs. Left panel exhibits a pattern typical from a sequential-suppression "scenario" while, on the right panel, an agreement with experimental CMS data is obtained within error bars, except for the most central collisions where our model lacks some $\Upsilon(1S)$ suppression. Comparing with our previous results~\cite{Gossiaux_Katz:2016}, this excess can be traced to be due to the larger 1S-survival -- thanks to P$\rightarrow$S transitions -- not completely compensated by the suppression of the feed-downs. We fear we cannot conclude on this point without including cold nuclear matter effects and considering AuAu and UU collisions at RHIC as well, what will be done in a future work.

\ack
We gratefully acknowledge the support from the TOGETHER project, R\'egion Pays de la Loire (France).

\section{References}


\begin{thebibliography}{9}
\bibitem{STAR:2014}
STAR Collaboration, Phys.Lett. B{\bf 735} (2014) 127-137, Erratum: Phys.Lett. B{\bf 743} (2015) 537-541.

\bibitem{CMS:2015}
CMS Collaboration, CMS PAS HIN-15-001 (2015).

\bibitem{ALICE:2014}
ALICE Collaboration, Phys. Lett. B{\bf 738} (2014) 361-372.

\bibitem{Gossiaux_Katz:2016}
P.B. Gossiaux and R. Katz, proceedings of Quark Matter 2015 conference [arXiv:1601.01443 [hep-ph]].

\bibitem{Katz:2015qja}
R. Katz and P.B. Gossiaux, Annals Phys. 368 (2016) 267-295, [arXiv:1504.08087 [quant-ph]].

\bibitem{Mocsy:2008}
A. Mocsy and P. Petreczky, Phys. Rev. D {\bf 77} (2008) 014501. 
\bibitem{Kaczmarek:2005}
O. Kaczmarek and F. Zantow,  Eur. Phys. J. C {\bf 43} (2005) 63. 

\bibitem{Andronic:2016}
A. Andronic et al., 
Eur.Phys.J. C{\bf 76} (2016) no.3, 107,
[arXiv:1506.03981 [nucl-ex]]. 

\bibitem{EPOS:2010}
 K. Werner, Iu. Karpenko, T. Pierog, M. Bleicher, K. Mikhailov,
   Phys.Rev. C {\bf 82} (2010) 044904.
  








 
\end{thebibliography}
\end{document}